\documentstyle[preprint,eqsecnum,aps,prb]{revtex}

\begin{document}
\draft
\title{Band gap and stability in the ternary intermetallic compounds\\
NiSnM (M = Ti, Zr, Hf):
A first principles study}
\author{Serdar \"{O}\u{g}\"{u}t and Karin M. Rabe}
\address{Yale University, Physics Department\\
P. O. Box 208120, New Haven, Connecticut, 06520-8120}
\date{\today}
\maketitle
\begin{abstract}

The structural stability and electronic properties of
the ternary intermetallic compounds NiSnM
(M = Ti, Zr, Hf) and the closely related Heusler compounds
Ni$_2$SnM are discussed
using the results of {\em ab initio} pseudopotential total energy
and band-structure calculations performed with a plane
wave basis set using the conjugate gradients algorithm.
The results characterize the lowest energy phase of
NiSnM compounds, with a SnM rocksalt structure sublattice, as
narrow gap semiconductors with indirect gaps near 0.5 eV,
while the Ni$_2$SnM compounds are
described as normal metals.  Two other
atomic arrangements for NiSnM in the MgAgAs structure type
result in energetically
unfavorable compounds which are metallic.
The gap formation in the lowest
energy structure of NiSnZr and relative
stability of the three atomic arrangements are investigated
within a tight-binding framework and by considering
the decompositions of each ternary compound into
a binary substructure plus a third element sublattice.
The stabilization of the lowest energy phase of NiSnZr is
found to be mainly due to the relative stability of the
SnZr rocksalt substructure, while the opening of the
gap induced by the addition of the symmetry-breaking Ni
sublattice makes a relatively minor contribution.
The results from the theoretical calculations for the NiSnM
compounds are compared with the existing experimental data.
{}From analysis of structural and chemical trends in
the NiSnM compounds, CoVSn is predicted to be
a new semiconducting intermetallic compound in
the MgAgAs structure type. Preliminary first principles
calculations suggest an indirect gap of 0.8 eV.

\end{abstract}


\narrowtext

\section{INTRODUCTION}

Recent discoveries of unconventional electronic properties
in multinary materials with complex crystal structures,
such as quasicrystals and high $T_c$ superconductors,
have stimulated a growing demand for
realistic theoretical studies of these systems.
A full understanding of the interrelationships
between composition, structure, and electronic properties
of these complex systems requires a detailed
analysis of individual materials at the microscopic level.
While the quantum mechanical modeling of solids through
first principles calculations has served as a very useful
tool to obtain the microscopic information needed for
structurally and chemically simple solids, there are
practical problems associated with complex systems, especially
with nonperiodic structures like quasicrystals.
The challenge, therefore, is to identify simpler systems
which can be fully studied with this theoretical tool, and
yet shed light on issues that are relevant to the physics of
more complex systems.

One issue that arises in the study of complex intermetallic
systems is the formation of a (pseudo)gap in
the electronic density of states (DOS) at the Fermi level
and the relation of this
feature to the stability of the particular structure.
For quasicrystals, the presence of an (almost) universal
suppression in the electronic DOS at the Fermi level has long
been established both experimentally\cite{Poon}
and from first principles band-structure calculations
for rational approximants.\cite{Fujiwara}
It is widely believed that this particular feature of
quasicrystals is related to their stability\cite{Hume1}
based on an analogy to the Hume Rothery rule\cite{Hume2}
for the stability of metallic alloys. The
role of a pseudogap in structural stability has also been
suggested by the experimental observation of very high
resistivities in structurally well ordered stable
quasicrystals.\cite{Exp,Pierce}
Motivated by these observations, there has been
recent interest in understanding the structural
energetics, stability and transport properties  of intermetallic
compounds with small unit cells exhibiting a (pseudo)gap in
their electronic spectrum.
However, relatively few examples of such systems are
known. CsAu is an extensively studied example\cite{CsAu}
with a band gap of 2.6 eV. The electronic
structure and transport properties of Al$_2$Ru
have recently been of considerable interest due to
its structural simplicity
and compositional similarity to quasicrystals.\cite{Pierce,Manh}
Another interesting class of intermetallic compounds
with a pseudogap at the Fermi level is the
skutterudite pnictides MA$_3$ (M = Co, Rh or Ir, and
A = P, As or Sb). These compounds exhibit unusual transport
properties making them likely candidates for
advanced thermoelectric materials.\cite{Singh}
The identification of additional examples of intermetallic
compounds with (pseudo)gaps is therefore a problem
of both fundamental and technological interest.

A few years ago, the ternary intermetallic compounds NiSnM
(M = Ti, Zr, Hf)
were reported to exhibit unusual transport
and optical properties suggesting that they may
have a gap at the Fermi level.\cite{Aliev1,Aliev2,Aliev3,Aliev4}
This observation establishes a possible
new class of semiconducting intermetallics, which we have
identified as good subjects for the theoretical study of
band gap formation in intermetallics and its relation
to structural stability.\cite{Ogut1}
These compounds crystallize in the MgAgAs
structure and are related to the metallic Heusler
compounds Ni$_2$SnM by the removal of an fcc Ni
sublattice.\cite{Villars} This simple crystal
structure makes them quite suitable for a first principles
approach\cite{Zunger} to the microscopic understanding of
their unusual optical and electronic properties.
While a complete understanding of these
properties would require incorporating the effects of
a high concentration of defects, calculations for the ideal
structures should provide a useful starting point for
further analysis. To understand the
origin of the semiconducting gap in this system,
one needs more than ``simple'' arguments.
{}From a tight-binding
analysis of the first principles bands and charge densities,
both the $pd$ hybridization on the SnM rocksalt
substructure and the symmetry breaking due
to the Ni sublattice are found to be
crucial factors in gapping the electronic spectrum.
Examination of total energies of ternary compounds and hypothetical
binary substructures, calculated from first
principles, allows us to conclude that the formation of
the gap is not directly associated with
structural stability in this system.
Finally, the use of first principles information
also assists us in identifying chemical trends for the prediction
of related semiconducting intermetallics.

The rest of the paper is organized as follows. In Section II,
we briefly discuss the technical details of the
{\em ab initio} pseudopotential calculations for the
NiSnM and Ni$_2$SnM compounds. In Section III, we first examine
their ground state structural properties and identify the correct
atomic arrangement for the NiSnM compounds. This is
followed by the results of band-structure and DOS
calculations
which characterize, within the local density
approximation (LDA), the Ni$_2$SnM
compounds as normal metals, and the NiSnM compounds as
narrow gap semiconductors with indirect gaps
near 0.5 eV. In Section IV, we discuss these results
in a tight-binding framework and by considering
decompositions of the ternary structures
into binary substructures plus a third element sublattice.
Particular attention is given to the origin
of the gap and structural stability of NiSnZr.
We compare our results with available experimental
measurements and, using the lessons from our
analysis, we predict a new semiconducting intermetallic compound.
Finally, our results are summarized in Section V.

\section{CALCULATIONAL PROCEDURE}

For the calculation of band structures and total energies,
we used the {\em ab initio} pseudopotential
method with a plane wave basis set and the conjugate gradients
algorithm.\cite{Payne1}
The general technical issues that arise
in the application of this method to intermetallic systems
containing transition metals have been discussed
in detail in Ref. \onlinecite{Ogut2}.
In this study, for Sn, Zr and Hf, we used the
scalar-relativistic pseudopotentials of Bachelet, Hamann,
and Schl\"{u}ter\cite{Bhs} (BHS).
For Ti and Ni, we constructed Hamann, Schl\"{u}ter, and
Chiang (HSC) pseudopotentials\cite{Hsc}
for the $s$ and $p$ orbitals, and optimized
pseudopotentials\cite{Rappe} (OPT) for the $d$ orbitals.
The exchange-correlation potential used in these
calculations is the functional of
Ceperley and Alder as parametrized by Perdew and
Zunger.\cite{Excorr}
Complete information concerning pseudopotential constuction
is given in Table I. These
pseudopotentials were put into separable form using
single or double projectors for each angular momentum
(Table I) with the $l=0$ component as the local
potential.\cite{Proj} In all cases, the absence of ghost
states and the transferability of pseudopotentials were
checked by applying the ghost theorem of
Gonze {\em et al.}\cite{Gonze} and
comparing the logarithmic derivatives for these
nonlocal pseudopotentials with their all-electron values
over a wide energy range.

All total energy and band-structure calculations were done
on IBM RS/6000 workstations using the conjugate gradients
program {\sc castep} 2.1,\cite{Payne2}
slightly modified for metallic systems.
The plane wave cutoff used in the calculations was
determined by requiring the kinetic energy in the
Fourier components of all atomic pseudo wave functions
to be less than 1 mRy. This cutoff was set
by the sharply peaked 3$d$ valence states of Ni to be 52 Ry,
allowing us to perform accurate calculations
with 2200-2800 plane waves depending on the size of the
unit cell. Due to the metallic nature of some of the
compounds investigated in this work, we used a
Fermi-Dirac broadening \cite{Gillan} of 50 meV
for {\bf k}-point sampling with
Monkhorst-Pack\cite{Monkhorst} grids
of $q = 8$ and $q= 10 $, resulting in 10 and
19 {\bf k} points
in the irreducible Brillouin zone (BZ), respectively.
Lattice constants and bulk moduli were obtained by fitting
to the Birch form, as described in Ref. \onlinecite{Ogut2}.

The band-structure calculations for the NiSnM and Ni$_2$SnM
compounds were performed with
self-consistent charge densities computed with a
Monkhorst-Pack grid of $q = 10$.
For DOS calculations, we first
used this charge density to evaluate the band structure
on a Monkhorst-Pack grid of $q= 20$ (110 {\bf k} points
in the irreducible BZ), and interpolated the eigenvalues
to a grid of $\approx$ 225~000 {\bf k} points in the
irreducible fcc BZ using the interpolation scheme of
Monkhorst and Pack\cite{Monkhorst}
based on a global BZ fit. We then applied the
Gilat-Raubenheimer method\cite{Gilat} to this fine mesh
in the irreducible BZ. To check the accuracy of this
interpolation, the calculated band structures along
certain high symmetry lines were compared with the
interpolated values, and the agreement was found to be
very good within a few percent.
Finally, for the band-structure plots and for use
in the tight-binding analysis, symmetry labels along
high symmetry lines were assigned
using projection operators for the
corresponding irreducible representations.

\section{RESULTS}

The ternary crystal structures included in this study are the
BiF$_3$ structure for the Ni$_2$SnM compounds and the
MgAgAs structure for the NiSnM compounds.\cite{Villars}
The crystal structure of Ni$_2$SnM
(space group $F$m\={3}$m$), shown in Fig. 1(a), can
be viewed as a rocksalt arrangement of Sn and M atoms with
Ni atoms at the center of each cube of Sn-M atoms.
The noncentrosymmetric crystal structure of NiSnM
(space group $F$\={4}$3m$), shown in Fig. 1(b),
can be obtained from that of Ni$_2$SnM
by eliminating one of the fcc Ni sublattices,
resulting in tetrahedral coordination of the Sn and
M atoms by Ni. We also considered
two additional distinct \mbox{MgAgAs} type crystal structures
obtained by interchanging the Ni, Sn and M sublattices.
These three different crystal structures can be distinguished
by identifying the doubly tetrahedrally coordinated element;
the relevant coordination environments are shown in Fig. 2(a).
We refer to these three structures as the $\alpha$, $\beta$,
and $\gamma$ phases depending on whether Ni, Sn, or M is
doubly tetrahedrally coordinated, respectively.\cite{Zunger}

First, we performed total energy calculations at several
lattice constants for all six
compounds Ni$_2$SnM and NiSnM (in the $\alpha$ phase) to determine
their ground state structural properties. The results are
summarized in Table II. As seen from this table, the calculated
equilibrium lattice constants are in very good agreement
with experiment, with slight underestimates typical of the LDA.
To our knowledge, no measurements of the bulk moduli
for these compounds have been reported in the literature;
the values in the table are therefore predictions for
the bulk moduli of the Ni$_2$SnM and NiSnM compounds.
We also performed total energy calculations for the $\alpha$,
$\beta$, and $\gamma$ phases of NiSnM. Figure 2(b) shows the
results of these calculations for M = Zr, and the $\alpha$
phase is seen to be the lowest energy structure (2.28 and
2.85 eV/unit cell below the $\beta$ and $\gamma$ phases,
respectively), consistent with reported structural determinations.
The experimental lattice constant for the NiSnZr compound
(6.11 \AA)\cite{Villars} is also closest to
the LDA lattice constant of the $\alpha$ phase.
This calculation therefore unambiguously determines the crystal
structure of the NiSnM compounds as the $\alpha$ phase.
The reasons behind the relative stability of the
$\alpha$, $\beta$, and $\gamma$
phases will  be discussed in Section IV.

Next, we performed band-structure and DOS calculations for both
Ni$_2$SnM and NiSnM compounds as a first step towards a
microscopic understanding of their electronic structures.
Figure 3 shows the calculated band structure along high
symmetry lines of the fcc BZ for Ni$_2$SnZr.
This exhibits
normal metallic behavior with bands crossing the Fermi
level along various directions, which results in a finite DOS
at the Fermi level, as shown in Fig. 4. Very similar
band structures and DOS were also obtained for Ni$_2$SnTi
and Ni$_2$SnHf. On the other hand,
band-structure calculations for the lowest energy
structure of NiSnM (the $\alpha$ phase) show that
the Fermi level does not intersect any bands along the
high symmetry directions, as can be seen in Fig. 5
for the case of NiSnZr. In fact,
DOS calculations show that for these compounds there is
a gap in the electronic DOS at the Fermi level (Fig. 6).
For all three compounds, this is an indirect
$\Gamma\rightarrow X$ gap near 0.5 eV, while the
direct gap, close to 1.1 eV, occurs at $X$ (Table III).
In contrast, band-structure calculations for the higher energy
$\beta$ and $\gamma$ phases of NiSnM show metallic behavior
(Fig. 7).
Therefore, these LDA calculations characterize
the NiSnM compounds in the $\alpha$ phase as narrow
gap semiconductors, while Ni$_2$SnM compounds
and the $\beta$ and $\gamma$ phases of NiSnM
are found to be normal metals.

\section{DISCUSSION}

Using the results of first principles total energy
and band-structure calculations, we can
develop a physical understanding of
the microscopic mechanisms in the formation of the
band gap in the $\alpha$ phase of NiSnM,
and of the differences between these
compounds and related systems, specifically the
$\beta$ and $\gamma$ phases and Ni$_2$SnM.
In Section IV A, we investigate the origin of the
band gap in a tight-binding framework. In Section IV B, we
consider the three phases of the ternary compounds NiSnM
using the decompositions into a binary substructure plus a third
element sublattice, and
show that these decompositions can be quite useful in
discussing the stability and electronic structures of ternary
compounds. After a brief discussion of the differences between
the metallic Heusler compounds Ni$_2$SnM and the semiconducting
NiSnM compounds in Section IV C, we examine the
available experimental
data on the NiSnM compounds from the point of view
of our theoretical calculations in Section IV D.
The relationships between band gap formation and
structural stability in intermetallic compounds are
considered in Section IV E, and
the applicability of these ideas to the
NiSnM compounds is examined. Finally, in Section IV F,
we use the lessons gained
from this analysis to predict a new semiconducting
intermetallic compound. We note that
for definiteness, the discussion
in these sections is for M = Zr, though the overall
features are the same for M = Ti and Hf.

\subsection{Origin of the band gap}

Our tight-binding analysis begins with identification of
the orbital characters
of the valence and conduction bands.
With a plane-wave basis set, this is accomplished
by examining
contour plots of the charge densities of
individual states at various {\bf k} points.
We first consider the bands well below the
Fermi level. The lowest-lying band in
Fig. 5 has predominantly Sn $s$ orbital character,
while the narrow bands above it have predominantly
Ni $d$ character.
This seems to suggest a picture in which bands arising
from the Sn $s$, Ni $d$,
and either Sn $p$ or $t_{2g}$ type Zr $d$ orbitals
are fully occupied, forming a closed shell system with
a gap to the next excited state.
We therefore examined the triply degenerate $\Gamma_{15}$
and the doubly degenerate $X_5$ states just below the
Fermi level, shown in Fig. 8(a).
The charge density plots in the (110) plane show
that the $\Gamma_{15}$ states have mainly Zr $d$
character [Fig. 8(c)], while the $X_5$ states
have mainly Sn $p$ character [Fig. 8(e)].
This change in orbital character shows that the
gap in NiSnM compounds cannot be due to a simple
closed shell picture, since in that case we
would expect the highest occupied band
to be derived from either Sn $p$ or Zr $d$ orbitals.

A detailed analysis of band behavior is therefore needed
to explain the formation of the gap.
We focus on the bands $\Gamma-\Delta-X$ as
capturing the essential features of the gap mechanisms.
Symmetry labels and orbital characters
allow the separation of bands into different
subspaces which can be considered one at a time.
We start with the highest occupied
valence band $\Delta_{3,4}$.
The doubly degenerate $\Delta_{3,4}$ conduction band from
$\Gamma_{15}$ to $X_5$  [Fig. 8(a)] has the same symmetry
label, but complementary orbital character, changing
from mainly Sn $p$ at $\Gamma$ [Fig. 8(b)] to
mainly Zr $d$ at $X$ [Fig. 8(d)].
To describe this pair of bands, we use a
tight-binding basis including the Sn $p_x$, $p_y$,
and Zr $d_{xz}$, $d_{yz}$ orbitals.\cite{Footnote1}
The behavior of these bands, shown with thick lines in Fig. 8(a),
can be explained as arising from dispersive Zr $d$ and Sn $p$
type bands which hybridize
along $\Delta$ and anticross. The reason that
this does not look like a familiar anticrossing of two bands
(where the bands approach and then slightly repel each other)
is that this $pd$ hybridization is large and
strongly ${\bf k}$-dependent, with the maximum
value halfway along $\Delta$.
Specifically, starting with nonzero interactions only for
Sn-Sn and Zr-Zr nearest neighbors in the fcc lattice,
the dispersion $E_{x,y}(\alpha)$ for
the Sn $p_x$ and $p_y$ orbitals is
\begin{equation}
E_{x,y}(\alpha)=\epsilon_p+2(pp\sigma+pp\pi)+
(2pp\sigma+6pp\pi)\cos \pi\alpha
\label{Ep}
\end{equation}
while the dispersion $E_{xz,yz}(\alpha)$
for the Zr $d_{xz}$ and $d_{yz}$ orbitals is
\begin{equation}
E_{xz,yz}(\alpha)=\epsilon_d+2(dd\pi+dd\delta)+
(3dd\sigma+2dd\pi+3dd\delta)\cos \pi\alpha
\label{Ed}
\end{equation}
where $\alpha$ runs from 0 (at $\Gamma$) to 1 (at $X$). In these
equations, $\epsilon_p$ and $\epsilon_d$ denote the onsite
orbital energies of the $p$ and $d$ orbitals, respectively,
and the definitions of the tight-binding
parameters are as in Ref. \onlinecite{Harrison}.
If the parameters are taken to have the usual
approximate relative magnitudes and signs
($dd\pi\approx-\frac{1}{2}dd\sigma$,
$dd\delta\approx 0$, $dd\sigma < 0$,
$pp\pi\approx-\frac{1}{4}pp\sigma$, and
$pp\sigma > 0$),\cite{Harrison} the $p$ type bands
disperse downward with a maximum at $\Gamma$, while the
$d$ type bands disperse upward with a minimum at $\Gamma$.
With a suitable choice
of the parameters, the two bands will cross.
However, including a nearest neighbor Sn-Zr
$pd$ interaction will result in
an anticrossing of these two bands, with
a matrix element $2pd\pi\sin \pi\alpha$ for this geometry,
which is largest halfway along $\Delta$.
This tight-binding parametrization reproduces
the first principles dispersion and is
consistent with the orbital character of the states obtained
from charge density plots.
This shows that the strong $pd$ hybridization in the
SnZr rocksalt substructure is a necessary ingredient in the formation
of the gap, for if it were zero, the $\Gamma_{15}$ states
below (above) the Fermi level would join with the $X_5$ states
above (below) the Fermi level resulting in a finite DOS at the
Fermi level. However, this hyridization is not in itself sufficient
for the gap formation in the ternary NiSnZr compound.

To complete the analysis of gap formation in the $\alpha$
phase of NiSnZr, we must consider the conduction band
minimum $X_3$ [Fig. 8(a)], which has mainly Zr $d_{xy}$
character.
Since the valence band maximum $\Gamma_{15}$
also has Zr $d_{xy}$ character, an unhybridized Zr $d_{xy}$
band, with $\Delta_1$ symmetry,  must anticross with
$\Delta_1$ bands derived from
other orbitals for the gap to form.
However, this
anticrossing cannot be explained by considering {\em only} the
SnZr substructure as in the case of $\Delta_{3,4}$ bands.
Due to the $O_h$ symmetry
of the rocksalt substructure, the matrix
elements between the Zr $d_{xy}$ band along
$\Delta$ and the bands derived from all
other Zr $d$ and Sn $p$ orbitals vanish,
resulting in zero anticrossing.
For the gap to form, it is necessary to include the
Ni $d$ orbitals, lowering the symmetry of the system
to $T_d$.
Although the Ni $d$ states lie well below the
Fermi level, the Ni $d_{xy}$ and $d_{3z^2-r^2}$
orbitals mediate an {\em indirect} interaction between the
Zr $d_{xy}$ band and the bands derived from the
Zr $d_{3z^2-r^2}$ and
Sn $p_z$ orbitals. These interactions lead to a nonzero
anticrossing of these bands, opening the gap at the Fermi
level. Therefore, the symmetry breaking produced by the third element
(Ni) sublattice is the essential additional ingredient in the
formation of the gap.

\subsection{Ternary compounds from binary substructures}

The analysis of the previous section showed that
the states near the Fermi level in the $\alpha$ phase
of NiSnZr are dominated by
the atomic orbitals of the SnZr rocksalt substructure.
This suggests an approach to the analysis of the
electronic structure and stability of the ternary
compound through the identification of a
key binary substructure.
In this approach, the essential features of electronic structure
and stability should already be evident in the
binary substructure, while addition of the third
element sublattice leads to recognizable
modifications such as (i) introduction of extra orbitals
hybridizing with the orbitals of the binary substructure,
(ii) change in the symmetry of the crystal structure, and
(iii) charge transfer between the binary substructure and
the third element sublattice, and other changes
in the self-consistent charge density.

We apply this approach to NiSnZr by computing the total energies
and band structures for the binary substructures appearing
in the decomposition of the $\alpha$, $\beta$ and $\gamma$ phases.
There are six distinct substructures: SnZr, NiZr and NiSn in the
rocksalt ($B1$) and zincblende ($B3$) structures.
The total energies of these six substructures
at the lattice constant of the $\alpha$ phase, with lists
of the corresponding NiSnZr phases, are given in Table IV.
For each of the three binary compositions, the
lower energy structure (between $B1$ and $B3$) is the one
contained in the decomposition of the $\alpha$ phase.
However, consideration of the magnitudes of
the $B1$-$B3$ energy differences singles
out the decomposition containing the SnZr substructure. The $B1$
structure of SnZr is 2.03 eV lower in energy than the $B3$
structure. This value is quite close to the energy difference between
the $\alpha$ phase, containing the $B1$ SnZr substructure,
and the $\beta$ and $\gamma$ phases, which contain the
$B3$ SnZr substructure (2.28 eV and 2.85 eV, respectively).
In contrast, the $B1$-$B3$ energy differences for NiZr and NiSn
are too small (0.26 eV and 0.23 eV per unit cell,
respectively) to be relevant to the stabilization of the $\alpha$
phase relative to the $\beta$ and $\gamma$ phases.
Therefore, the low energy of the NiSnZr compound in the $\alpha$
phase can be mainly attributed to the stability of
the SnZr $B1$ substructure.

There also appear to be useful relationships between the
band structures of the ternary compounds and those of binary
substructures.
For example, the $\alpha$ phase of NiSnZr, where the SnZr
substructure is rocksalt, has a gap while the $\beta$ and $\gamma$
phases, with SnZr in a zincblende arrangement, are both metallic.
We investigate these relationships in more detail by examining the
band structures of the six substructures (Fig. 9) and comparing them
with the band structures of the ternary $\alpha$ [Fig. 8(a)], $\beta$
[Fig. 7(a)] and $\gamma$ [Fig. 7(b)] phases.
In the $B3$ phases of NiSn and NiZr,
the crystal field splittings of the Ni and Zr $d$ states
at $\Gamma$ are noticeably larger than in the $B1$ phases. The
reason is that the point symmetry is lowered from $O_h$ in the
$B1$ structure to $T_d$ in the $B3$ structure,
resulting in a mutual repulsion of $t_{2g}$ type
$d$ and $p$ orbitals in the latter.
Comparison of the $B1$ with the $B3$ band structures for the
three compositions shows a striking difference between NiZr
and NiSn, on the one hand, and SnZr, on the other hand,
in the position of the Fermi level and dispersion
of the nearby bands. The very small change
in the position of the Fermi levels for the two structures
of NiZr and NiSn is not surprising, because the Ni $d$ orbitals
are rather low-lying, and do not substantially affect states around
the Fermi level in these compounds. In contrast, the arrangement
of the bands and position of the Fermi level changes noticeably
from SnZr $B1$ to SnZr $B3$. This occurs because
Zr $d$ and Sn $p$ orbitals are partially occupied
orbitals involved in bonding, making
the band dispersions and the positions of the
Fermi level quite sensitive to the atomic arrangement.

We can also compare the band structures of the
three phases of the ternary
compound with the band structures of the $B1$
and $B3$ structures of SnZr that appear in their decompositions.
For example, comparison of Fig. 8(a) and the band structure of SnZr
in the $B1$ structure (Fig. 9) shows that the low-lying Sn $s$ band,
the conduction bands well above the Fermi level, and the
bands arising from the Sn $p_x$, $p_y$ and Zr $d_{xz}$, $d_{yz}$
hybridizations (labelled by $\Delta_5$ in Fig. 9) look quite
similar to those in Fig. 8(a). On the other hand, the $\Delta_1$
band in Fig. 9, arising from the doubly degenerate $\Gamma_{12}$
state and ending up as the $X_{2}^{\prime}$ state at $X$ is
substantially modified by the introduction of the Ni
sublattice. In the ternary compound,
this band anticrosses due to the presence of Ni $d$
orbitals and reduced point group symmetry.
Closer examination shows that
the introduction of the Ni $d$ orbital greatly increases
the crystal field splitting of the Zr $d$ states at $\Gamma$,
with the Fermi level in the $\alpha$ phase of NiSnZr
placed just inside this splitting. Thus,
within the substructure decomposition approach,
we can account for the formation of the gap.
The band structure of NiSnZr in the $\gamma$ phase [Fig. 7(b)]
also has similarities with the band structure of SnZr
in the $B3$ structure (Fig. 9). Since the Sn $p-$Zr $d$ interaction
pushes the $t_{2g}$ type Zr $d$ states down in energy in the $B3$
structure, the introduction of Ni $d$ orbitals results in the
formation of a $d$-band complex below the Fermi level, while
the downward dispersing bands well above the Fermi level
are not substantially affected by this.
On the other hand, the relationship between the
band structure of the $\beta$ phase of NiSnZr [Fig. 7(a)]
with the band structure of its $B3$ SnZr substructure
is much more complicated, because
introducing the Ni $d$ orbitals changes the mixing of the $p$ and $d$
orbitals to a greater extent due to double tetrahedral coordination
of Sn. In this case, it is not possible to treat the effects of the
Ni $d$ orbitals perturbatively.

\subsection{Comparison of Ni$_2$SnM with NiSnM}

The Ni$_2$SnM compounds can be produced
from the $\alpha$ phase of NiSnM
by adding a Ni sublattice, which changes the system from a
narrow gap semiconductor to a metal.
The effect of the additional sublattice responsible for this change
in electronic structure is the raising of the
noncentrosymmetric $T_d$ symmetry to the
centrosymmetric $O_h$ point group symmetry, with
consequent changes in the symmetry properties of atomic-orbital
Bloch functions.
In particular, if $\phi$ denotes any of the
five $d$ or $s$ orbitals of Ni, the functions
that transform according to the irreducible
representations of the point group of
any high symmetry point in the BZ have the form
$\frac{1}{\sqrt{2}}(\phi^{(1)}\mp\phi^{(2)})$,
where $\phi^{(i)}$ ($i = 1, 2$) denotes the $\phi$ orbital
of the $i^{\scriptsize\mbox{th}}$ Ni atom in the
unit cell.
Also, any orbital
that transforms according to $\Delta_1$ ($X_3$)
in the noncentrosymmetric
compound transforms in Ni$_2$SnM according to either
$\Delta_1$ or $\Delta^{\prime}_2$ ($X_4$ or $X^{\prime}_2$).
Therefore, the $\Delta_1$ and $\Delta^{\prime}_2$ bands
near the Fermi level (Fig. 3)
do not anticross as they do in NiSnM, where they both
carry the $\Delta_1$ label.
As a result, the $\Delta_1$ band cuts the Fermi level,
producing the metallic behavior.
Finally, in our
calculations we do not see
any evidence of a vacancy band\cite{Aliev4} proposed to
explain the experimental data for NiSnM. This
is not too surprising for the perfectly clean samples
on which our calculations were based, because unlike
the case of heavy doped semiconductors, the vacancies
in this system are very well ordered (in an fcc lattice)
and have a high concentration.

\subsection{Comparison with experiment}

What light do these calculations shed on the experimentally
observed electronic and optical properties of the NiSnM
compounds? First of all, the very high
low-temperature resistivities with a simple
exponential temperature dependence observed
in the NiSnM compounds are consistent with the
the appearance of a semiconducting gap in the
calculated band structures.
However, specific heat measurements show a small
but finite DOS at the Fermi level.\cite{Aliev4}
Furthermore, the calculated gaps
are all near 0.5 eV.
These are more than twice as large as
the gap parameters
0.12, 0.19, and 0.22 eV
for Ti-, Zr-, and Hf-based compounds, respectively,
extracted from resistivity measurements\cite{Aliev3} and
features in the optical reflectivity $R(\omega)$.\cite{Aliev2}
While LDA errors (either underestimates
or overestimates) in the values of the band gaps are expected,
a more serious problem
which could give rise to these discrepancies is
poor sample quality,
particularly substitutional disorder.
In NiSnZr, there is direct evidence from x-ray diffraction for
Sn-Zr substitutional disorder at the 10-30\% level.\cite{Aliev1}
The degree of substitutional disorder seems to be correlated with
changes in the electronic properties.
One example of this is that
the DOS at the Fermi level appears to decrease as the degree
of disorder decreases.
In particular, the
zero temperature resistivity increases dramatically upon
annealing the samples.\cite{Aliev1} Also,
the DOS at the Fermi level from specific heat
measurements\cite{Aliev4} is lowest for the Ti-based
compound, which has
the smallest disorder in the position of the Ti and Sn
atoms, indicating that a complete gap may be expected to open
in a very clean sample.
To improve our understanding of the
effect of the Sn-Zr substitutional disorder on the size of the gap,
we performed total energy and band-structure calculations
for Ni(Sn$_{1-x}$Zr$_x$)(Sn$_x$Zr$_{1-x}$) for
a few values of $x$ between 0 and 0.5 in the virtual
crystal approximation. At around $x = 0.15$,
the gap closes to form a semimetal, with the valence band maximum
(conduction band minimum) moving up (down) in energy and
touching the Fermi level.
Therefore, to achieve experimental gap values comparable
to those calculated for the ideal crystals, special attention to
sample preparation is needed.
In such well-ordered samples, reflectivity measurements would have
to be extended to higher frequencies than those used in previous
studies (in Ref. \onlinecite{Aliev2} $\omega<$ 5000 $cm^{-1}$ )
to see the onset of absorption.

\subsection{Band gap and structural stability}

Direct relations between the
formation of a (pseudo)gap and the stability of the
structure have been proposed in several contexts,
including the Hume Rothery rule,\cite{Hume1,Hume2}
Jahn-Teller effects in molecules, \cite{Jahn} and
the ``coloring'' patterns of atoms over the sites
of a given lattice.\cite{Burdett} If
the formation of a gap in the DOS is an important
factor in the structural stability of intermetallic compounds,
the expectation is that semiconducting intermetallics would be
fairly common, while, as we have already discussed, they are
quite rare. Therefore, it is useful
to examine this issue in the present case.

For NiSnM compounds in their $\alpha$,
$\beta$, and $\gamma$ phases, it is true that
the structure with the gap has the lowest energy,
and there is a correlation between the DOS at the Fermi
level and the structural energies of the three phases.
However, the present calculations suggest that the formation of
the gap is not directly responsible for the structural stability.
Instead, the stability can be understood from the substructure
decomposition approach.
As discussed in Sec. IV B, the energy of the SnZr $B1$
structure is 2.03 eV per unit cell lower than that of the
$B3$ structure. From Fig. 9, it is clear that neither the
$B1$ nor the $B3$ SnZr structures have a gap.
The ternary phases are obtained by incorporating the
Ni sublattice into the binary SnZr substructures.
This incorporation produces a gap
in the $\alpha$ phase, as a result of symmetry breaking, but
leads to only a small change in the relative
energy (from 2.03 eV per unit cell to energy differences of
2.28 eV and 2.85 eV per unit cell between the
$\alpha$ and the $\beta$ and $\gamma$ phases, respectively).
Therefore, the stabilization of the lowest energy phase is primarily
due to the relative stability of the binary SnZr $B1$
substructure, while the contribution
to stabilization associated with the incorporation
of the third element (Ni) sublattice, which results in
the formation of the gap, is relatively minor.
This conclusion is not inconsistent with the relations between gap
formation and stability observed elsewhere. In the Hume-Rothery
rule, a gap in the DOS establishes the limit of
stability of a particular alloy structure as a
function of electron count, rather than an enhanced stability
specifically of the semiconducting phase.
In the Jahn-Teller and coloring analyses, the correlation between
structural stability and gap formation is identified in classes of
structures which are otherwise extremely similar (related by small
structural relaxations in the Jahn-Teller case and possessing
identical nearest neighbor environments for the
cations in the coloring problem). However, in the NiSnM compounds,
the $\alpha$, $\beta$ and $\gamma$ phases are related by sublattice
interchanges, not small displacements, resulting in different
nearest neighbor environments in each phase.
It is therefore reasonable that the substructure decomposition
approach, with the identification of the low-energy
rocksalt SnZr substructure, should be a more useful indicator of
structural stability than gap formation for these compounds.

\subsection{Considerations for new semiconducting intermetallics}

{}From these results,
the simple chemical and structural trends
in gap formation can be identified and used to
assist in the prediction of new semiconducting intermetallics.
We have already seen that a closed shell rule with 18
electrons is too simple. The ternary nature of the compounds,
particularly the electronegativity difference between the
constituents of the rocksalt substructure,
and the structural
arrangement, particularly the double tetrahedral coordination
of the late transition metal,
seem to be important factors, since
the hypothetical compounds NiSn$_2$ and NiZr$_2$, and
the $\beta$ and $\gamma$ phases of NiSnM compounds are
metallic.\cite{Footnote2}
On the other hand, the empirical rule that
the total number of $d$ electrons from the two
transition metals in their ground states should be 10
works for the NiSnM compounds as well as for the compounds
formed by the column
substitutions of Ni with Pd and Pt.\cite{Aliev3}
It is also necessary that
the transition metal that is doubly tetrahedrally
coordinated should have a significantly larger number of
$d$ electrons in the atomic ground state, so that
the bands originating from these $d$ orbitals
will lie deep below the Fermi level, and not
destroy the $pd$ hybridizations by mixing significantly
with the $d$ orbitals of the other transition metal.
CoVSn in the MgAgAs structure with Co doubly tetrahedrally
coordinated by Sn and V satisfies these rules.
First-principles band structure calculations
confirm that CoVSn has an $L\rightarrow X$ gap of 0.8 eV
at the Fermi level (Fig. 10).
This compound
is not reported to exist in the intermetallic compound
database. However, there is a good chance that this might be a stable
compound based on the analogy with the $\alpha$ phase of NiSnM
compounds.  Therefore, we predict that if
this compound can successfully be prepared in this structure,
it will be a semiconducting intermetallic.

\section{CONCLUSIONS}

In this paper, we used the pseudopotential total energy
method to investigate the stability and electronic
structure of the ternary intermetallic compounds
NiSnM (M = Ti, Zr, Hf). Our calculations
revealed an indirect semiconducting gap near
0.5 eV for all three compounds
in the lowest energy $\alpha$ phase,
while the inversion symmetric Ni$_2$SnM compounds
and the higher energy $\beta$ and $\gamma$ phases
of NiSnM were found to be metallic.
{}From a tight-binding
analysis of the first-principles bands and charge densities,
both the $pd$ hybridization on the SnZr rocksalt substructure
and the symmetry breaking due to the Ni sublattice were found to be
crucial factors in gapping the electronic spectrum of
$\alpha$-NiSnZr.
By considering all binary substructures which appear
in the decomposition of the three phases of NiSnZr,
we concluded that the stabilization of the lowest energy
$\alpha$ phase is primarily
due to the relative stability of the binary SnZr $B1$
substructure, while the contribution
to stabilization associated with the incorporation
of the third element (Ni) sublattice, which results in
the formation of the gap, is relatively minor.
Analysis of the band structures for the narrow
gap semiconductor NiSnZr and the metallic
Ni$_2$SnZr showed that the raising of the point group symmetry
from the noncentrosymmetric $T_d$
to the centrosymmetric $O_h$ group
by the addition of the Ni sublattice is responsible
for this difference in the electronic properties.
Our theoretical findings are consistent with experimental
observations, if poor
sample quality and substitutional disorder in
the samples are taken into account.
Finally, we predicted that the ternary compound
CoVSn in the MgAgAs structure with an SnV
rocksalt substructure will form a semiconducting intermetallic,
if it can be prepared successfully in this structure.

\acknowledgments

We would like to thank J. C. Phillips, L. F. Mattheiss,
W. A. Harrison, J. K. Burdett, J. Rodgers,
U. V. Waghmare, and R. B. Phillips for
many useful discussions. We would also like to thank M. C. Payne
for the use of {\sc castep} 2.1.
This work was supported by NSF Grant No.
DMR-9057442. In addition, K. M. R acknowledges
the support of the Clare Boothe Luce Fund and the Alfred
P. Sloan Foundation.

\begin{figure}
\caption{The crystal structures of (a) Ni$_2$SnM; (b) NiSnM
compounds.}
\end{figure}

\begin{figure}
\caption{(a) Three possible coordination environments corresponding
to the $\alpha$ (bottom), $\beta$ (middle), and $\gamma$ (top)
phases of NiSnM;
(b) Total energy versus the lattice constant for the $\alpha$,
$\beta$, and $\gamma$ phases of NiSnZr. The energies shown
in this graph are with respect to the minimum energy corresponding
to the LDA lattice constant of the $\alpha$ phase.}
\end{figure}

\begin{figure}
\caption{The energy bands of Ni$_2$SnZr along the high symmetry
lines of the fcc Brillouin zone. The calculations were performed
at the experimental lattice constant of 6.27 \AA. The dotted line
is the Fermi level. The symmetry labels are shown only for states
that are discussed in Section IV C.}
\end{figure}

\begin{figure}
\caption{Electronic density of states
for Ni$_2$SnZr. The calculations were performed at the experimental
lattice constant of 6.27 \AA. The dotted line is the
the Fermi level.}
\end{figure}

\begin{figure}
\caption{The energy bands of NiSnZr along the high symmetry
lines of the fcc Brillouin zone. The calculations were performed
at the experimental lattice constant of 6.11 \AA. The dotted line
is the Fermi level.}
\end{figure}

\begin{figure}
\caption{Electronic density of states
for NiSnZr. The calculations were performed at the experimental
lattice constant of 6.11 \AA. The dotted line is the
the Fermi level.}
\end{figure}

\begin{figure}
\caption{The energy bands of NiSnZr in the (a) $\beta$;
(b) $\gamma$ phase along $\Delta$. The calculations were
performed at a lattice constant of 6.11 \AA. The dotted
lines represent the position of the Fermi level.}
\end{figure}

\begin{figure}
\caption{(a) The energy bands of NiSnZr along $\Delta$.
The thick curves are the doubly
degenerate (due to time reversal symmetry)
$\Delta_{3,4}$ bands mentioned in the text;
(b)-(c) Charge density contour plots for the triply degenerate
$\Gamma_{15}$ states; (d)-(e) Charge density contour plots for
the doubly degenerate $X_5$ states. The circles filled with
$\times$ marks and thick dots represent the positions of
the Sn and Zr atoms, respectively.}
\end{figure}

\begin{figure}
\caption{The energy bands for the SnZr, NiZr, and NiSn binary compounds
in the $B1$ and $B3$ structures along $\Delta$. All calculations
were performed at a fixed lattice constant of 6.11 \AA.
The dotted lines represent the position of the Fermi level.}
\end{figure}

\begin{figure}
\caption{Band structure of CoVSn in the MgAgAs structure along
$L-\Lambda-\Gamma-\Delta-X$. The dotted line is the Fermi level.
The calculations were performed at a lattice constant of 5.9 \AA.}
\end{figure}

\mediumtext
\begin{table}
\caption{Parameters and methods used in the construction of the
pseudopotentials. For OPT, $q_{c}$ and N in the last two columns
refer to the cutoff above which the kinetic energy is minimized
and the number of Bessel functions used to expand the pseudo
wave functions, respectively.\protect\cite{Rappe}}
\begin{tabular}{lcccccc}
Element & Configuration & Type & Projector\tablenote{The
$l = 0$ component was chosen as the local pseudopotential.}
& $r_{c}$ (a.u) & $q_{c}$ (Ry)& N \\
\tableline
Ni&4s$^{0.75}$ 4p$^{0.25}$ 3d$^8$ & s, p HSC & p, single &
1.2, 1.2 & & \\
&&d, OPT& d, single& 2.0 &7.15&10\\
Ti&4s$^{0.75}$ 4p$^{0.25}$ 3d$^2$ & s, p HSC & p, single &
1.3, 1.4 & & \\
&&d, OPT& d, single& 1.9 & 6.80 &4\\
Sn&5s$^1$ 5p$^{0.5}$ 5d$^{0.5}$ &BHS& p, double &&&\\
&&&d, double&&&\\
Zr&5s$^{0.75}$ 5p$^{0.25}$ 4d$^2$ &BHS& p, double &&&\\
&&&d,double&&&\\
Hf&6s$^{0.75}$ 6p$^{0.25}$ 5d$^2$ &BHS& p, single &&&\\
&&&d,double&&&\\
\end{tabular}
\end{table}
\narrowtext

\begin{table}
\caption{The experimental and calculated structural properties of the
Ni$_2$SnM and NiSnM compounds.}
\begin{tabular}{lccc}
Compound & a$_{exp.}$ (\AA) \tablenote{From
Ref. \protect\onlinecite{Villars} and Fig. 5 in
Ref. \protect\onlinecite{Aliev3}.}
& a$_{calc.}$ (\AA) & B
(Mbar)\tablenote{The values in parentheses are the bulk
moduli calculated at the LDA lattice constant.} \\
\tableline
Ni$_2$SnTi & 6.09 & 6.00 & 1.64 (2.14) \\
Ni$_2$SnZr & 6.27 & 6.24 & 1.49 (1.59) \\
Ni$_2$SnHf & 6.24 & 6.13 & 1.36 (1.72) \\
NiSnTi & 5.92 & 5.814 & 1.11 (1.43) \\
NiSnZr & 6.11 & 6.07 & 1.19 (1.29) \\
NiSnHf & 6.066 & 5.95 & 1.10 (1.41) \\
\end{tabular}
\end{table}
\begin{table}
\caption{LDA band gaps for NiSnM compounds. The indirect gap
occurs along $\Gamma$ to $X$, while the direct gap occurs
at $X$.}
\begin{tabular}{lcc}
Compound & Indirect gap (eV) & Direct gap (eV) \\
\tableline
NiSnTi & 0.51 & 1.20\\
NiSnZr & 0.51 & 1.05\\
NiSnHf & 0.48 & 1.16\\
\end{tabular}
\end{table}
\begin{table}
\caption{Total energies for six different binary compounds
which appear in the decomposition of the $\alpha$, $\beta$,
and $\gamma$ phases of NiSnZr into $B1$ and $B3$ structures.
The Greek letters in the
parentheses under the second column give the phase of NiSnZr
which will be obtained by the addition of the third type of
element into either the tetrahedral or the octahedral hole
of the fcc unit cell. All calculations were performed at a fixed
lattice constant of 6.11 \AA.}
\begin{tabular}{lcc}
Compound & Structure & Total Energy (eV) \\
\tableline
SnZr & $B1$ ($\alpha$) & -180.200 \\
SnZr & $B3$ ($\beta$,$\gamma$) & -178.172 \\
NiZr & $B1$ ($\beta$) & -997.115 \\
NiZr & $B3$ ($\alpha$,$\gamma$) & -997.375 \\
NiSn & $B1$ ($\gamma$) & -1011.445 \\
NiSn & $B3$ ($\alpha$,$\beta$) & -1011.671 \\
\end{tabular}
\end{table}

\begin{references}
\bibitem{Footnote1} It should be noted that in this approximate
analysis, the Ni $d_{xz}$ and $d_{yz}$ orbitals which also
transform according to the same irreducible representations
along $\Delta$ are neglected. This is quite safe,
because these orbitals are rather deep-lying, and
do not have an appreciable mixing with either of the bands in the
vicinity of the Fermi level, as can be checked from charge
density plots. We also note that the functions
that transform according to the $\Delta_3$ and $\Delta_4$ irreducible
representations are not actually the Bloch sums
of $p_x$, $p_y$, $d_{xz}$, and $d_{yz}$ orbitals, but rather the
linear combinations $\frac{1}{\sqrt{2}}(p_x\mp p_y)$ and
$\frac{1}{\sqrt{2}}(d_{xz}\mp d_{yz})$. However, this does not
change the dispersion relations given by Eqs. (\protect\ref{Ep})
and (\protect\ref{Ed}), nor does it change the form of the
coupling between the Sn $p$ and Zr $d$ orbitals.
\bibitem{Footnote2} We should note that since the band overlap
in the $\beta$ phase is quite small, self-energy corrections
may also make this phase semiconducting, though with a
much smaller band gap compared to the $\alpha$ phase.
\end{references}
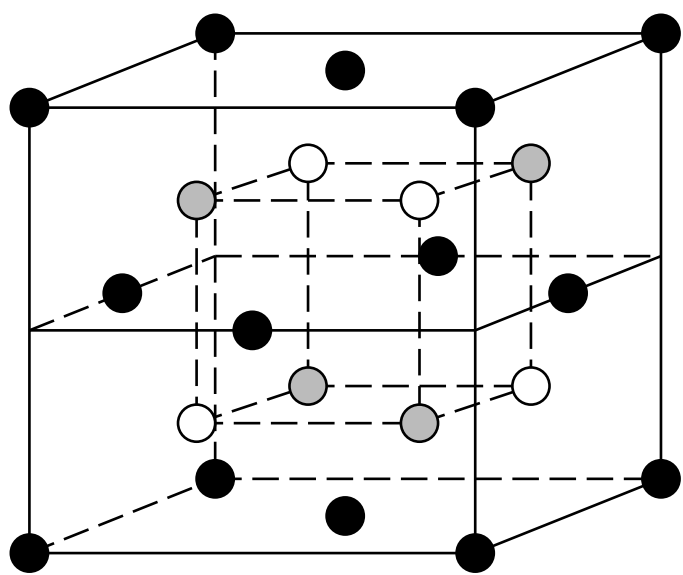
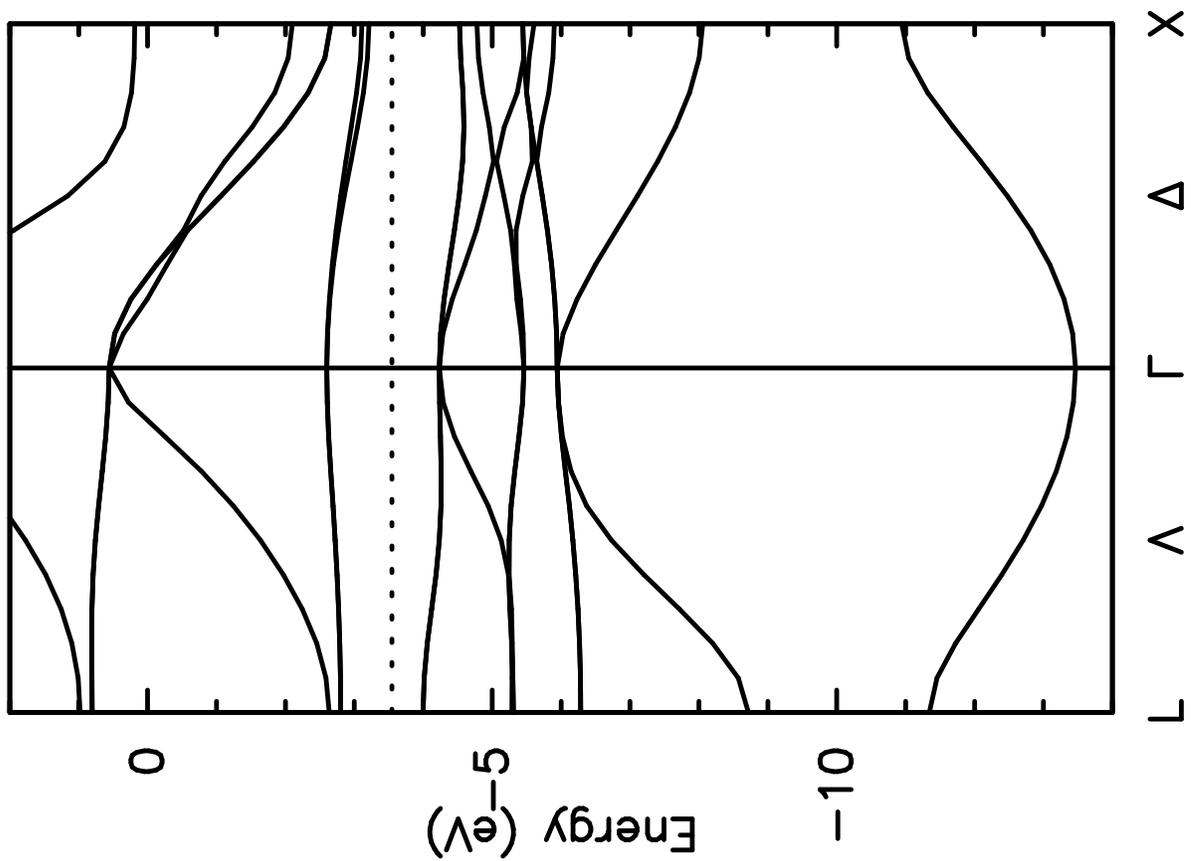
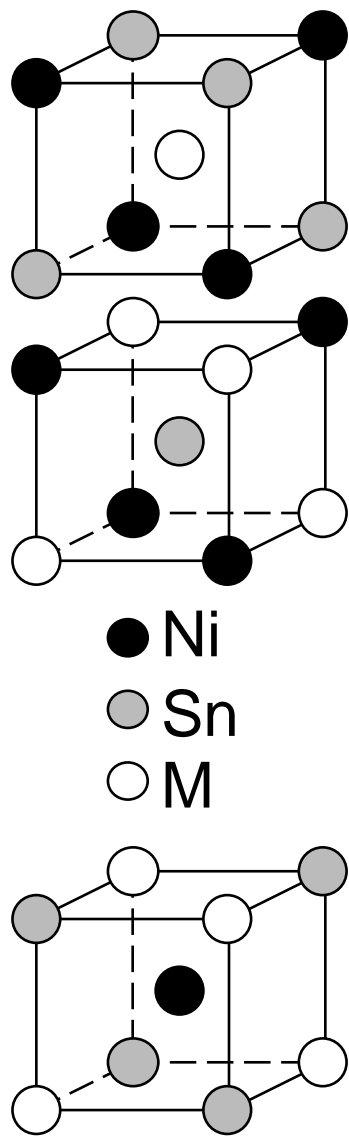
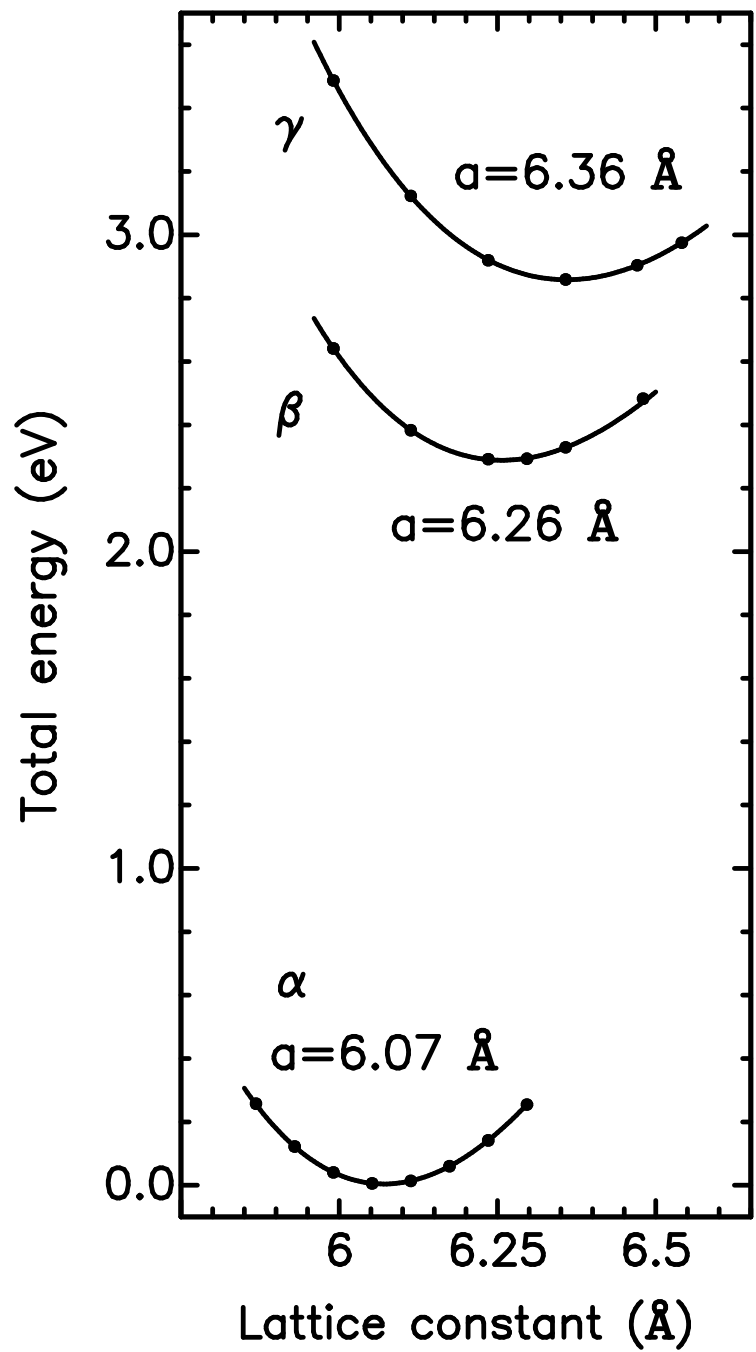
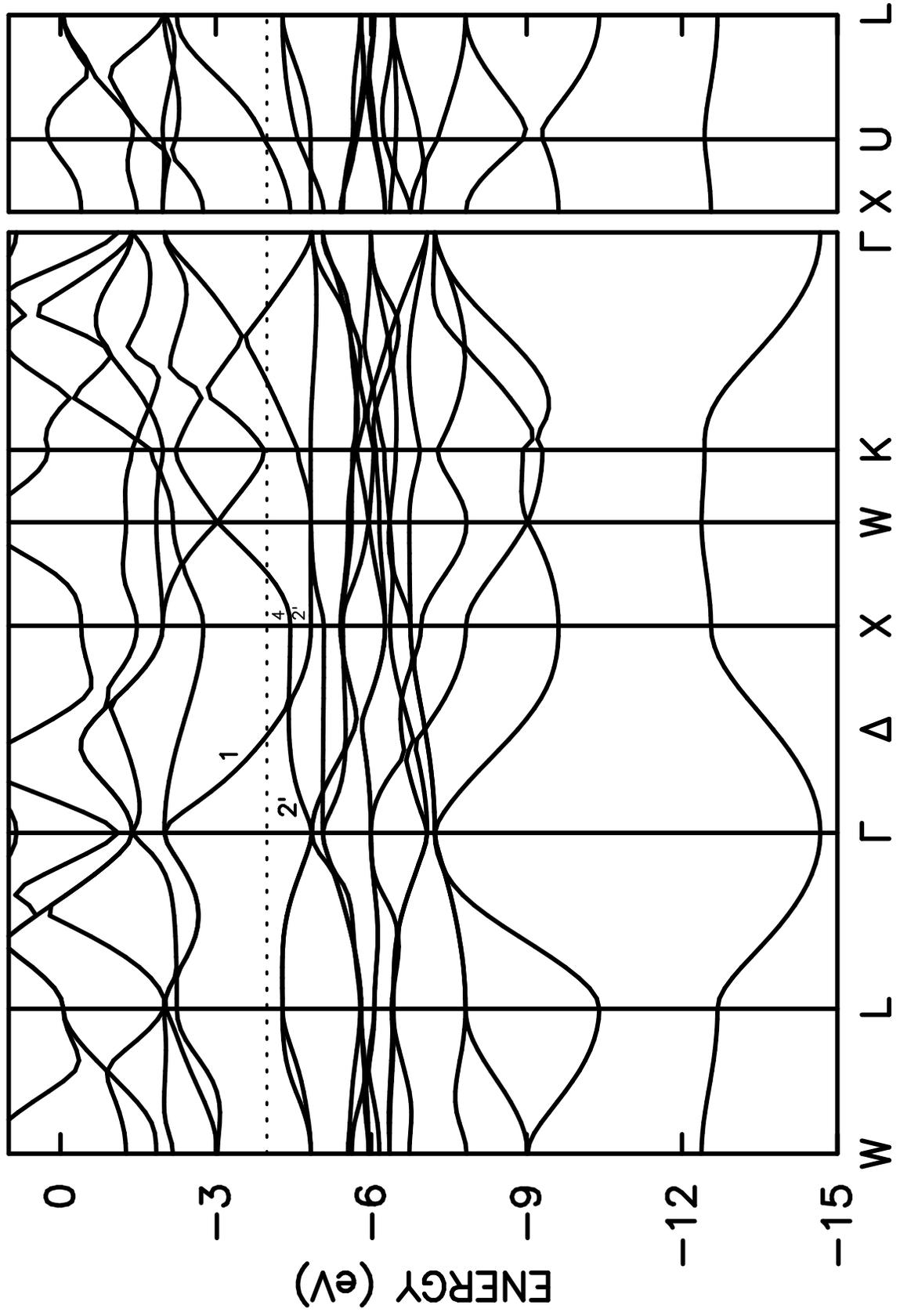
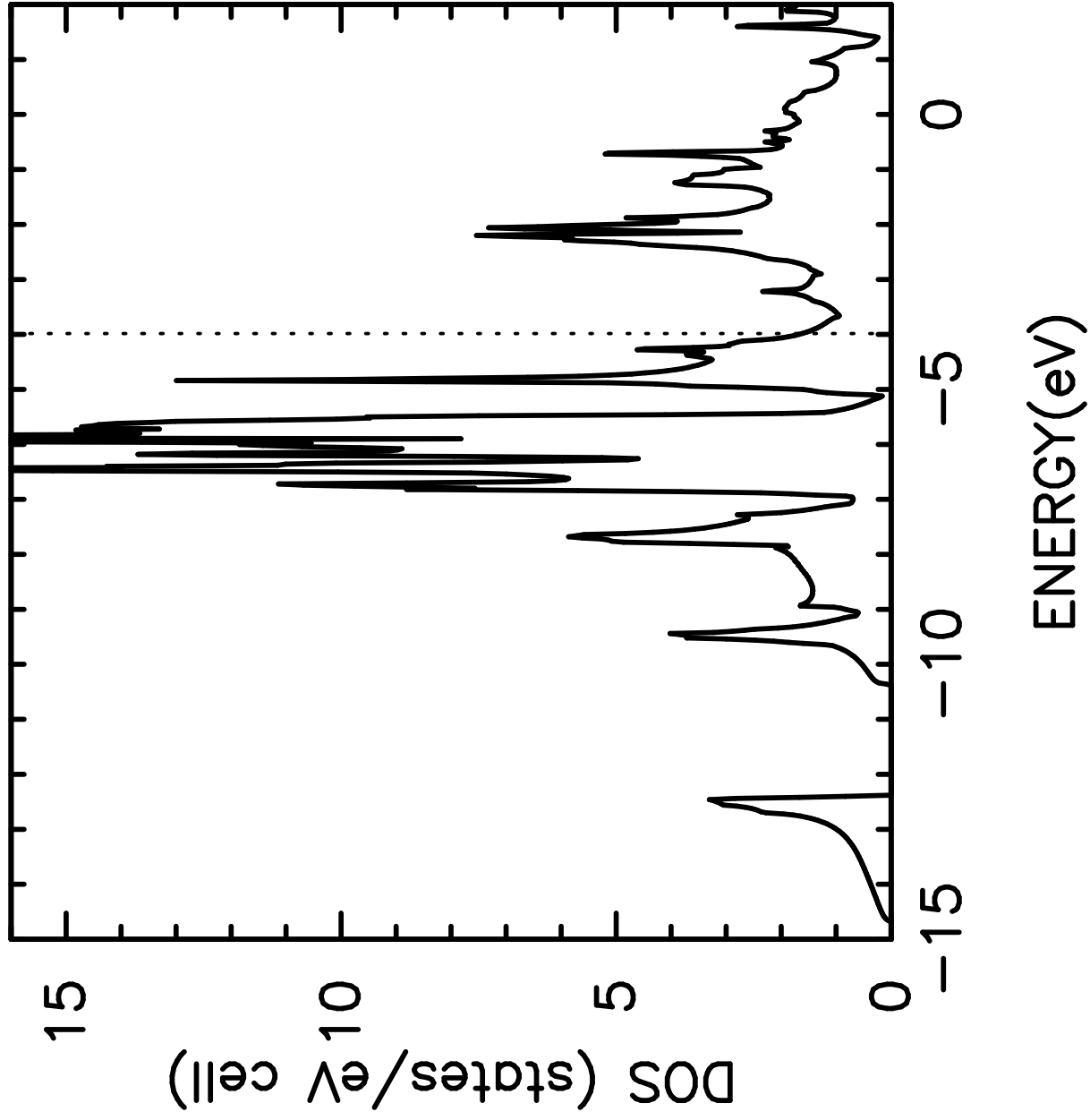
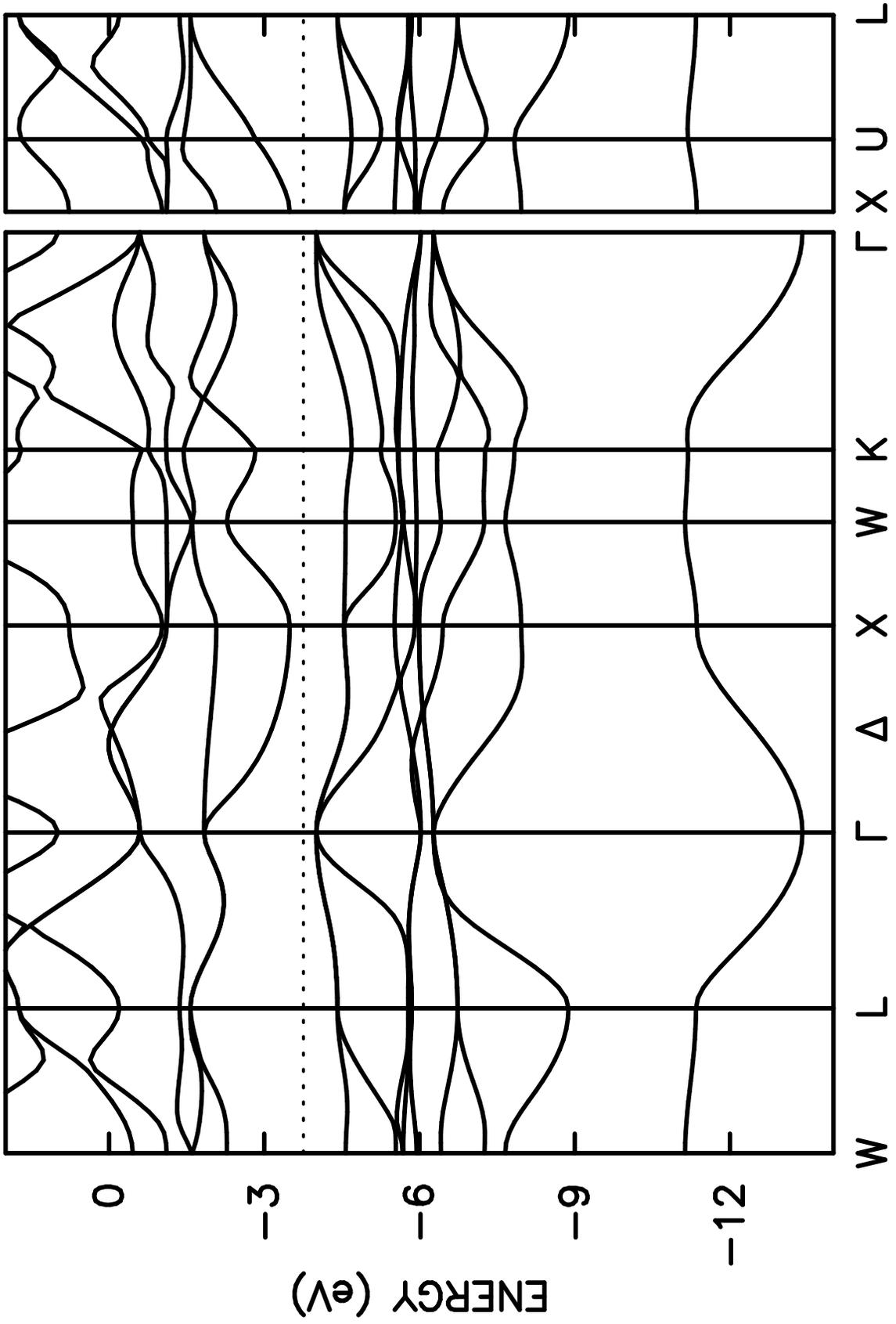
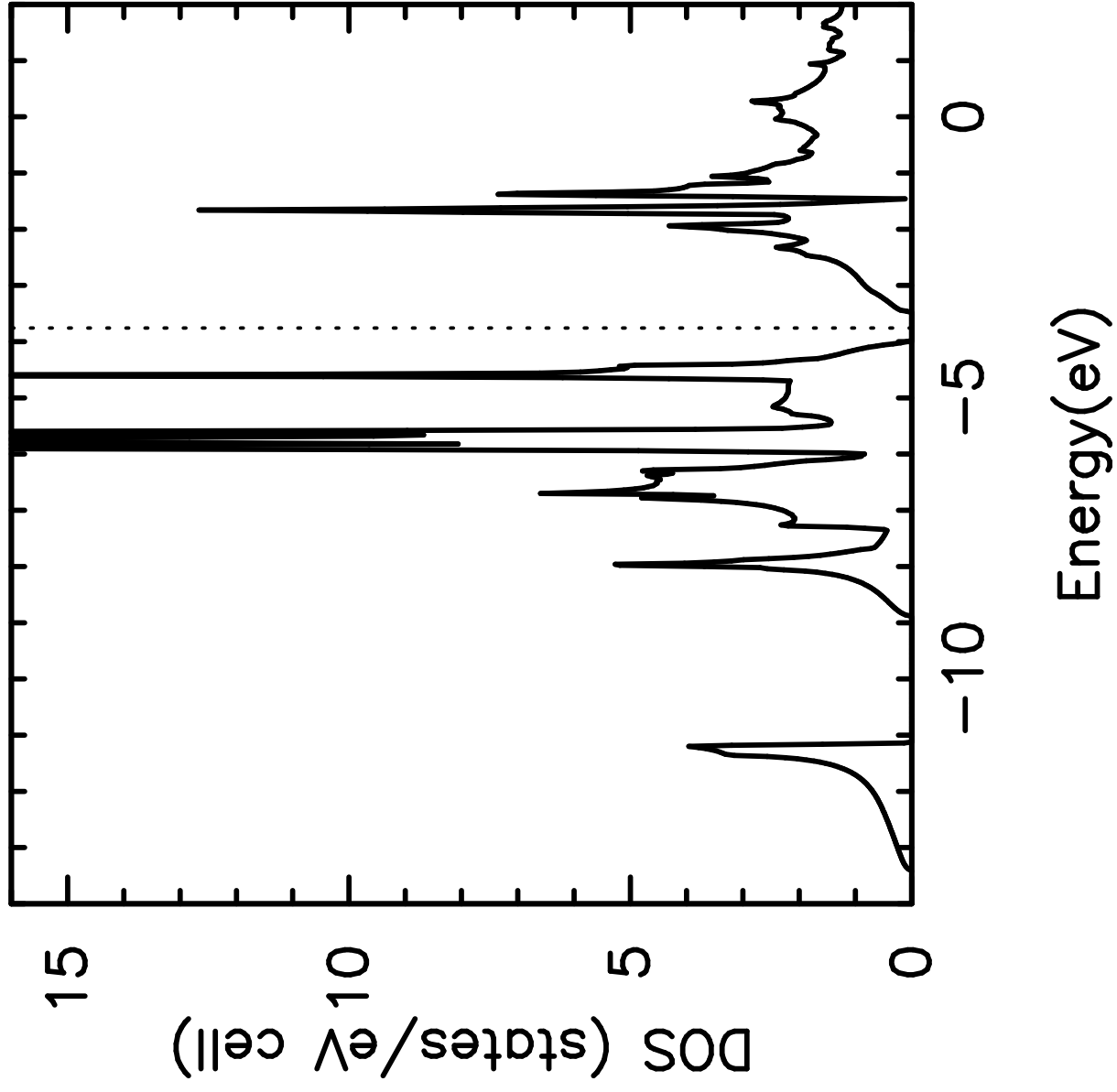
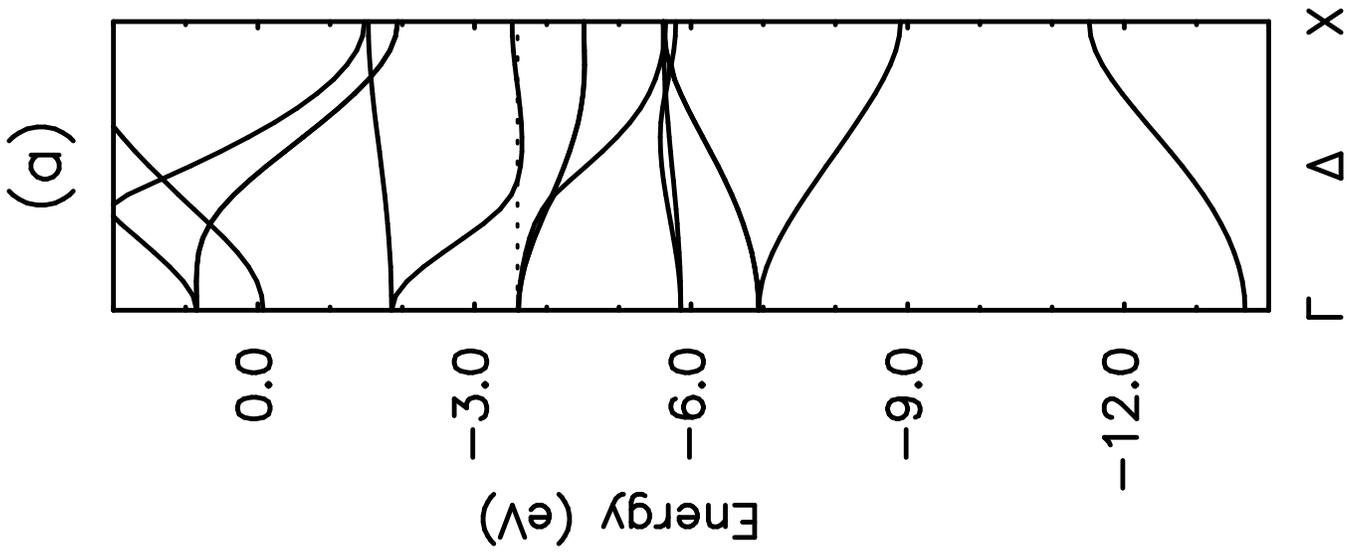
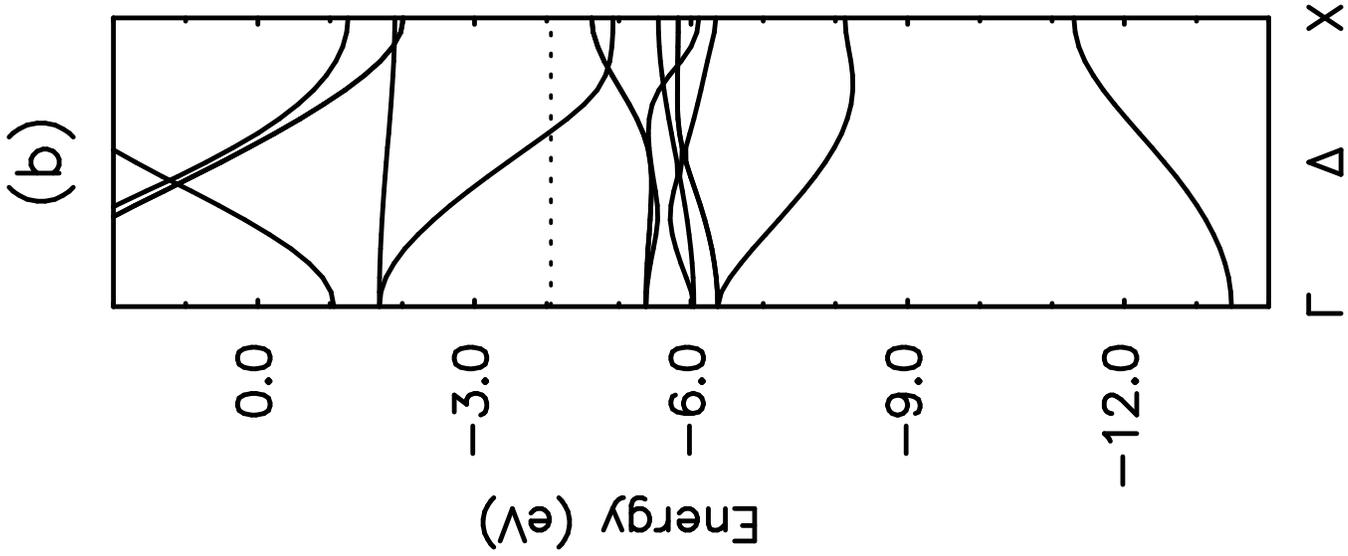
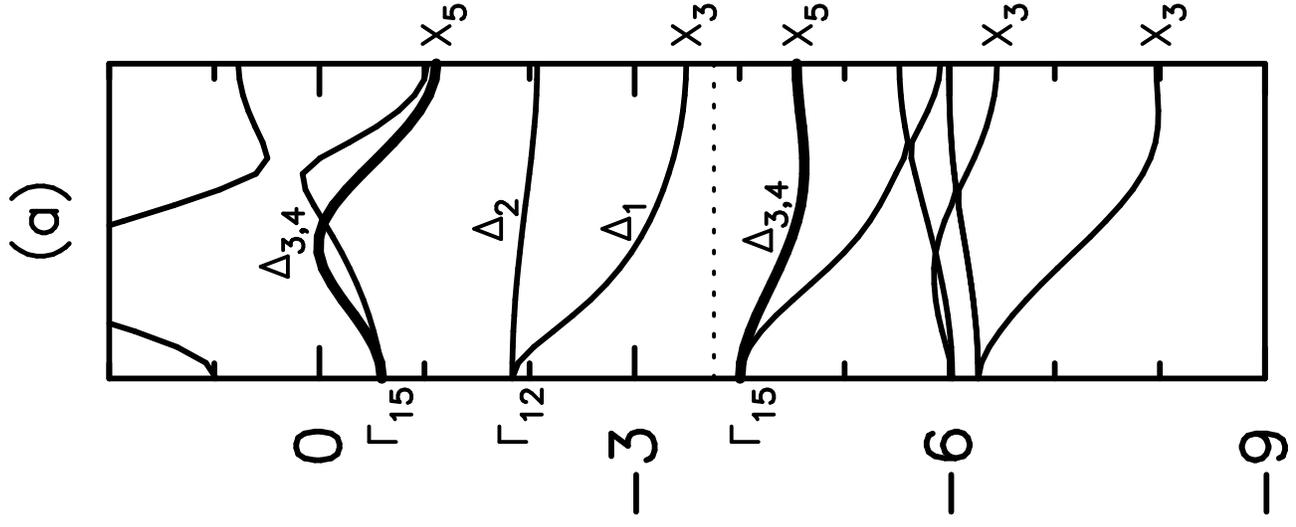
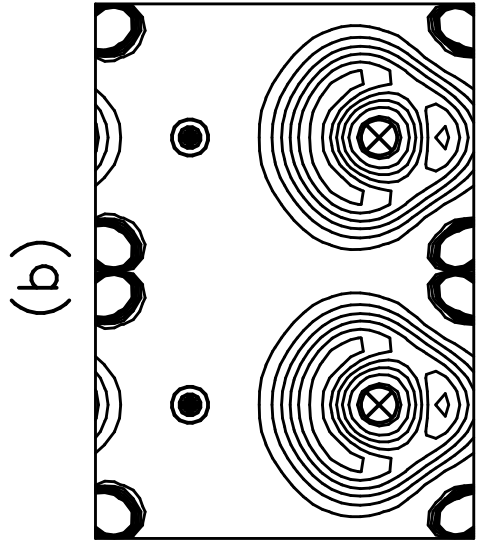
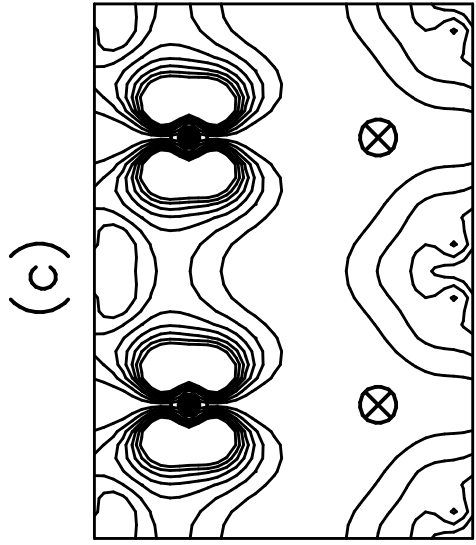
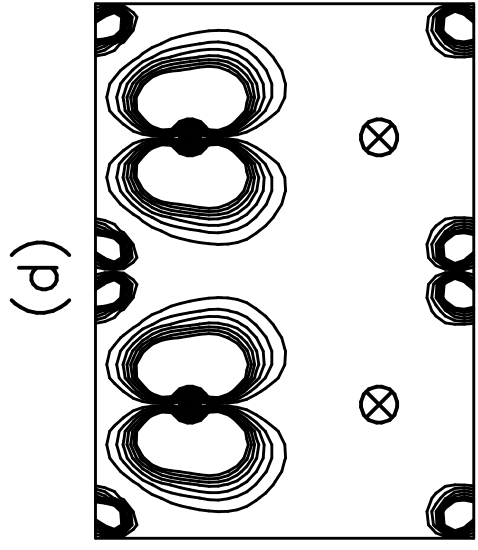
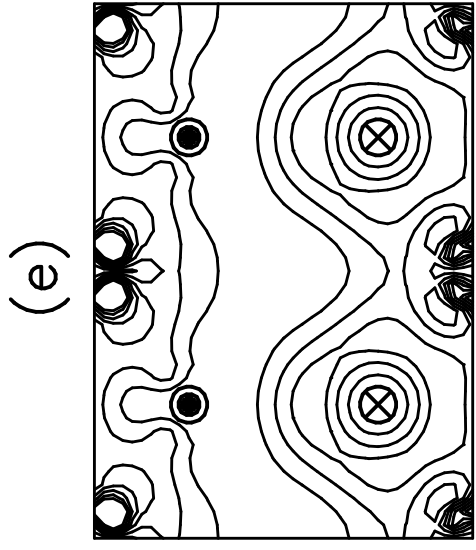
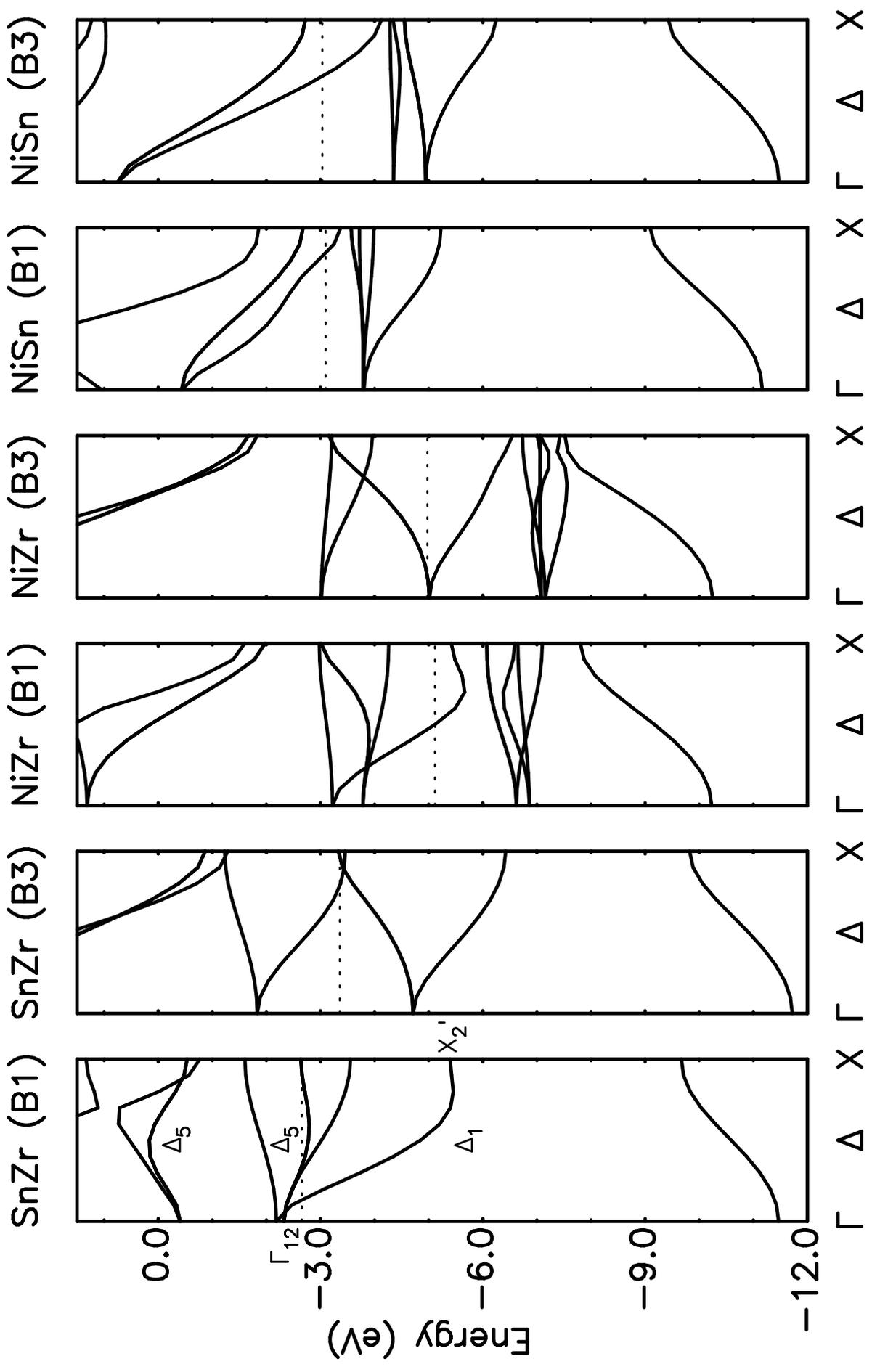
\end{document}